\documentclass[hidelinks,prd,eqsecnum,amsmath,amssymb,nofootinbib,12pt,superscriptaddress]{revtex4-1}

\usepackage{dcolumn}
\usepackage{bm}
\usepackage{xcolor}
\usepackage[applemac]{inputenc}
\usepackage[spanish,english]{babel}
\usepackage{amsmath,amssymb,amsfonts,latexsym,cancel,amsthm,hyperref,bbm}
\usepackage{graphicx}
\usepackage{color}
\usepackage{soul}
\usepackage{ulem}
\usepackage{slashed}
\usepackage{braket}
\usepackage[mathscr]{euscript}
\usepackage{physics}
\usepackage{mathrsfs}
\usepackage{pgfplots}
\usepackage{float}
\usepackage{xfrac}

\allowdisplaybreaks

\newcommand{\la}{\langle}
\newcommand{\ra}{\rangle}
\newcommand{\w}{\omega}

\newcommand{\be}{\begin{equation}}
\newcommand{\ee}{\end{equation}}
\newcommand{\bea}{\begin{eqnarray}}
\newcommand{\eea}{\end{eqnarray}}
\newcommand{\bes}{\begin{subequations}}
\newcommand{\ees}{\end{subequations}}

\begin{document}

    \title{\bf Quantum Effects in 3+1 Schwarzschild-de Sitter Spacetime: Properties of the Hadamard Function}
    \author{Ian M. Newsome}\email{newsim18@wfu.edu}
    \affiliation{Department of Physics, Wake Forest University, Winston-Salem, NC, 27109, USA}

    \author{Silvia Pla}\email{silvia.pla\_garcia@tum.de}
    \affiliation{Theoretical Particle Physics and Cosmology, King's College, London, WC2R 2LS, UK}
    \affiliation{Physik-Department, Technische Universit\"at M\"nchen, James-Franck-Str., 85748 Garching, Germany.}

    \author{Paul R. Anderson}\email{anderson@wfu.edu}
    \affiliation{Department of Physics, Wake Forest University, Winston-Salem, NC, 27109, USA}

\begin{abstract}

    In a four-dimensional Schwarzschild-de Sitter background, the spherically symmetric $(\ell=0)$ contribution to the Hadamard two-point correlation function is computed for a massless minimally-coupled scalar field in the Unruh state. Consideration is given to spacetime points located between the black hole and cosmological horizons. Previously it was found in two dimensions at late times for spatially separated points that the Hadamard function exhibits unbounded linear growth in time, with a rate of growth proportional to the sum of the black hole and cosmological surface gravities. Here it is shown numerically that this instability persists in four dimensions, but with a modification of the two-dimensional result due to scattering effects associated with the scalar field modes. An analytic approximation is derived for the growth rate in four dimensions and, in the limit that the black hole vanishes, is found to be equivalent to the rate of growth for the Hadamard function found previously for de Sitter space in cosmological coordinates.

\end{abstract}

    \date{\today}
    \maketitle

\newpage

\section{Introduction}

    Schwarzschild-de Sitter spacetime  (SdS) is an exact solution to the vacuum Einstein equations in the presence of a positive cosmological constant. It describes an eternal black hole immersed in an expanding universe and contains both a black hole and cosmological horizon. The Unruh state~\cite{Unruh} is the natural state for a quantum field in an eternal black hole spacetime to describe the late-time effects of black hole evaporation associated with a black hole that forms from collapse. Its generalization to SdS has been given in~\cite{MarkovicUnruh,TadakiTakagi}. In particular, for SdS one can associate Hawking radiation \cite{SHawking} with each horizon. For a given horizon, the temperature is $T_h = \kappa_h / 2\pi$, where $\kappa_h$ is the magnitude of the surface gravity of the horizon~\cite{GibbonsHawking}.

    It is interesting to study the behavior of the quantum field modes which comprise the Unruh state. For a massless minimally-coupled scalar field in two dimensions (2D), there is no scattering and, in the usual static coordinates, the mode functions for the Unruh state approach constant values at late times for fixed values of the spatial coordinate. In contrast, for a massive minimally-coupled scalar field in 2D SdS it was found~\cite{AndersonShohrehScofield} that the corresponding mode functions for the Unruh state vanish at late times for fixed values of the spatial coordinate. This was also found generically for a 2D static asymptotically flat black hole spacetime when a delta function effective potential is present in the mode equation. In both cases one can write the modes for the Unruh state in terms of packets of modes for the Boulware state (see e.g. \cite{AndersonShohrehScofield}). The difference between the late time behaviors of the Unruh state modes was shown to be correlated with the fact that, for a massless minimally-coupled scalar field in 2D, there are no scattering effects and therefore the infrared divergence present in the Boulware state modes, originating from normalization, persists. In contrast, for the case of a massive minimally-coupled scalar field in 2D SdS or a massless minimally-coupled scalar field in an asymptotically flat 2D eternal black hole spacetime with a delta function potential, scattering effects remove the infrared divergences in the Boulware modes.
    
    The behavior of the symmetric, or Hadamard, two-point function has been investigated~\cite{AndersonTraschen} for a massless minimally coupled scalar field in the Unruh state in 2D for SdS and other 2D spacetimes containing black hole and/or cosmological horizons. It was found in the static patch exterior to the black hole horizon and/or inside the cosmological horizon that late-time, unbounded linear growth occurs if the points are separated along a coincident time hypersurface in coordinates natural to the region. The rate of growth for 2D SdS was found to be
\be
    R_{2D} \equiv \frac{\partial}{\partial t}G^{(1)}(t,r;t,r') = \frac{1}{2 \pi} (\kappa_b + \kappa_c) \quad, \label{R-2D}
\ee
    where $\kappa_b$ and $\kappa_c$ are the surface gravities associated with the black hole and cosmological horizons, respectively. It was found in~\cite{AndersonShohrehScofield} for asymptotically flat eternal black hole spacetimes that no such growth in time occurs if a delta function effective potential is included in the mode equation. This appears to be related to the removal of infrared divergences in the Boulware modes by scattering effects.

    In this paper, the question of whether the above effects found in 2D SdS for a massless minimally-coupled scalar field in the Unruh state persist in four dimensions (4D) is addressed. Consideration is given to the spherically symmetric modes, whose gray-body factors do not vanish in the zero-frequency limit~\cite{Bradyetal, KantiGrainBarrau}. In this case scattering effects do not remove the infrared divergences in the Boulware modes~\cite{AndersonFabbriBalbinot}. In contrast, for $\ell >0$ the gray-body factors vanish in the zero-frequency limit~\cite{Crispinoetal,KantiPappasPappas} and therefore it is extremely likely that the infrared divergences in the Boulware modes will also be removed. In parallel with the previous 2D SdS calculations \cite{AndersonTraschen,AndersonShohrehScofield}, spacetime points located between the black hole and cosmological horizons will be considered.

    In Sec.~II, a review of SdS spacetime is given with a formulation of the quantum states and mode equation required for computation of the Hadamard function. In Sec.~III, numerical results for the 4D Unruh state modes and the Hadamard function are presented. An analytic approximation for the growth rate associated with the 4D Hadamard function is derived in Sec.~IV, including its predictions and comparisons between the 4D numerical and 2D analytic results. A summary and discussion of the results is given in Sec.~V.

\section{Quantum States and Mode Decomposition in \\ 3+1 Schwarzschild-de Sitter}
\label{states}

\begin{figure}[h]
   \centering
   \includegraphics[scale=0.36]{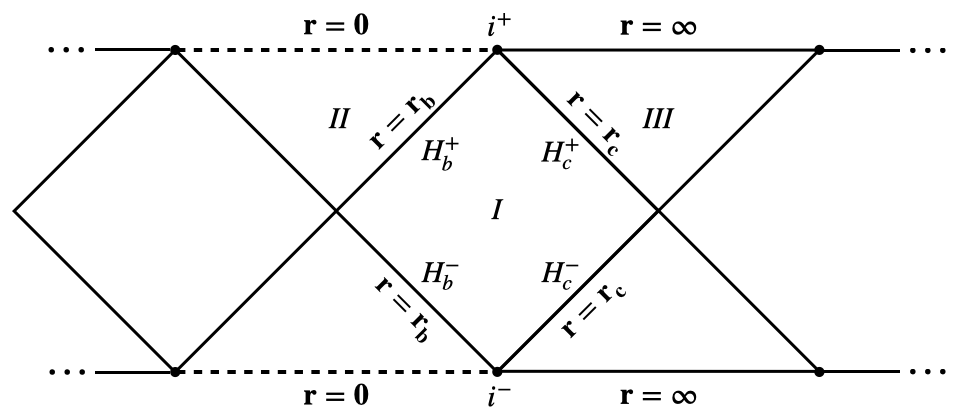}
   \caption{Penrose diagram depicting Schwarzschild-de Sitter spacetime. The past black hole and cosmological horizons are denoted $H^-_{(b,c)}$ while the future horizons are denoted $H^+_{(b,c)}$. The static patch is indicated by region $I$, whereas the black hole interior and cosmological far-field region are indicated by $II$ and $III$ respectively.}
   \label{SdS}
\end{figure}

    The metric for SdS in terms of the usual static coordinates is
\be
    ds^2 = - f(r) dt^2 + \frac{1}{f(r)}dr^2 + r^2 \left( d\theta^2 + \sin^2{\theta}d\phi^2 \right) \quad , \label{4DSdS1}
\ee
    with
\be
    f(r) = 1 - \frac{2M}{r} - H^2 r^2 = -\frac{H^2}{r}\big( r - r_b \big)\big( r - r_c \big)\big(r + r_b + r_c \big) \label{metricfunc} \quad .
\ee
    Here $M$ is the black hole mass, $H^2 = \Lambda/3$ with $\Lambda$ a positive cosmological constant, and the black hole and cosmological horizon radii are $r_b$ and $r_c$ respectively. Throughout, the units $\hbar=c=G=1$ are used.  A Penrose diagram for the SdS spacetime is given in Fig.~\ref{SdS}. From the relation $\eqref{metricfunc}$, one can show that~\cite{AndersonTraschen}
\be
    r_c = -\frac{r_b}{2} + \frac{1}{H}\sqrt{1-\frac{3H^2 r_b^2}{4}} \quad .
\ee
    If $r_b \to 0$, then $r_c \to 1/H$.  The black hole and cosmological horizons merge when $r_b=1/\sqrt{3}H$.

    The general form of the Hadamard function is 
\be
    G^{(1)}(x,x') = \langle \{ \phi(x) , \phi(x') \} \rangle \label{Hadamard1} \quad ,
\ee
    where the scalar field $\phi$ satisfies the equation
\be
    \Box \phi(x) = 0 \quad . \label{box-phi}
\ee
   The modes associated with the Unruh state are defined on the Cauchy surface consisting of the union of the past black hole and cosmological horizons $H^-_b$ and $H^-_c$.  Two sets of modes denoted by $p^{b}_{\omega \ell m}$ and $p^{c}_{\omega \ell m}$ comprise the Unruh state. On  $H^-_b$
\bes
    \begin{flalign}
        p^b_{\omega \ell m}(x) &= \frac{1}{\sqrt{4 \pi \omega}} \frac{e^{-i \omega U_b}}{r_b} Y_{\ell m}(\theta , \phi) \quad , \\
        p^c_{\omega \ell m}(x) &= 0 \quad ,
    \end{flalign}
\ees
while on $H^-_c$
\bes
    \begin{flalign}
        p^b_{\omega \ell m}(x) &= 0 \quad , \\
        p^c_{\omega \ell m}(x) &= \frac{1}{\sqrt{4 \pi \omega}} \frac{e^{-i \omega V_c}}{r_c} Y_{\ell m}(\theta , \phi)   \quad .
    \end{flalign}  \label{Kruskal1} 
\ees
    Here $U_b$ and $V_c$ are the Kruskal coordinates associated with $H^-_{b}$ and $H^-_{c}$, respectively. Hereafter $p^{(b,c)}_{\omega \ell m}$ are referred to as Kruskal modes. In relation to the null coordinates $u=t-r_*$ and $v=t+r_*$, with $dr_*=dr/f$, the quantities $U_b$ and $V_c$ are
\bes
    \be \label{eq:kkUb}
        U_b = -\frac{1}{\kappa_b}e^{-\kappa_b u} \quad , \quad r>r_b \quad ,
    \ee
    \be\label{eq:kkVc}
        V_c = -\frac{1}{\kappa_c}e^{-\kappa_c v} \quad , \quad r<r_c \quad .
    \ee
\ees
    As in~\cite{AndersonTraschen}, the surface gravity associated with a particular horizon $r=r_h$ is defined to be $2\kappa_h=|f'(r_h)|$ with the result that
\bes
    \begin{flalign}
        \kappa_b &= \frac{H^2}{2 r_b}(r_c-r_b)(2r_b+r_c) \quad , \\
        \kappa_c &= \frac{H^2}{2 r_c}(r_c-r_b)(r_b+2r_c) \quad , \\
        \kappa_N &= \frac{H^2}{2(r_b+r_c)}(r_b+2r_c)(2r_b+r_c) \quad .
    \end{flalign} \label{kappas}
\ees
   The quantity $\kappa_N$ denotes the surface gravity associated with the negative root of $\eqref{metricfunc}$, namely $r_N=-(r_b+r_c)$. An explicit form for $r_*$ is \cite{AndersonTraschen}
\begin{flalign}
    r_*(r) = \frac{1}{2\kappa_b}\ln{\left\{ \frac{|r-r_b|}{r_c-r_b} \right\}} &- \frac{1}{2\kappa_c}\ln{\left\{ \frac{|r-r_c|}{r_c-r_b} \right\}} + \frac{1}{2\kappa_N}\ln{\left\{ \frac{r+r_c+r_b}{r_c+2r_b} \right\}} \nonumber \\
    &- \frac{r_c}{4 r_b\kappa_b}\ln{\left\{ \frac{2 r_c+r_b}{r_c+2r_b} \right\}} - \frac{r_b r_c}{2(r_c-r_b)}\ln{\left\{ \frac{r_b}{r_c} \right\}} \quad .
\end{flalign}

    The scalar field for the Unruh state is
\be
    \phi(x) = \sum_{\ell=0}^\infty \sum_{m=-\ell}^\ell \int_0^\infty d\omega \bigg[ a_{\omega \ell m}^b \, p_{\omega \ell m}^b(x) + a_{\omega \ell m}^{b \, \dagger} \, p_{\omega \ell m}^{b \, *}(x) + a_{\omega \ell m}^c \, p_{\omega \ell m}^c(x) + a_{\omega \ell m}^{c \, \dagger} \, p_{\omega \ell m}^{c \, *}(x) \bigg] \label{scalarfield} \quad ,
\ee
    where $[a_{\omega \ell m}^h,a_{\omega' \ell' m'}^{\dagger \, h'}]=\delta(\omega-\omega') \, \delta_{\ell,\ell'} \, \delta_{m,m'} \, \delta_{h,h'}$ with $h=b,c$. It follows that the Hadamard function $\eqref{Hadamard1}$ for the Unruh state is
\be
    G^{(1)}(x,x') = 2 \sum_{\ell=0}^\infty \sum_{m=-\ell}^\ell \int_0^\infty d\omega \, \mathrm{Re}\bigg\{ p_{\omega \ell m}^b(x) \, p_{\omega \ell m}^{b \, *}(x') + p_{\omega \ell m}^c(x) \, p_{\omega \ell m}^{c \, *}(x') \bigg\} \label{G1kruskal} \quad .
\ee
    
    The mode  equation $\eqref{box-phi}$ is not separable in either set of Kruskal coordinates. However, in terms of the static coordinates in $\eqref{4DSdS1}$, it is separable and a complete set of solutions can be obtained in terms of the modes which comprise the Boulware\footnote{Here the natural generalization of the Boulware state from that in Schwarzschild spacetime~\cite{Boulware} is used.} state. On $H^{-}_b$
\bes
    \begin{flalign}
        h^{b}_{\omega \ell m}(x) &= \frac{1}{\sqrt{4 \pi \omega}} \frac{e^{-i \omega u}}{r_b} Y_{\ell m}(\theta, \phi)  \quad , \\
        h^c_{\omega \ell m}(x) &= 0 \quad ,
    \end{flalign}
\ees
    and on $H^{-}_c$
\bes
    \begin{flalign} 
        h^{b}_{\omega \ell m}(x) &= 0 \quad , \\
        h^{c}_{\omega \ell m}(x) &= \frac{1}{\sqrt{4 \pi \omega}} \frac{e^{-i \omega v}}{r_c} Y_{\ell m}(\theta, \phi) \quad .
    \end{flalign} \label{Boulware1}
\ees
    The Kruskal modes $p^{(b,c)}_{\omega \ell m}$ can be expanded in terms of the Boulware modes $h^{(b,c)}_{\omega' \ell' m'}$ using a Bogolubov transformation with the result
\be
    p_{\omega \ell m}^{(b,c)}(x) = \sum_{\ell'=0}^\infty \sum_{m'=-\ell'}^{\ell'}\int_0^\infty d\omega' \bigg[\alpha_{\omega \ell m \omega' \ell' m'}^{(b,c)} \, h_{\omega' \ell' m'}^{(b,c)}(x) + \beta_{\omega \ell m \omega' \ell' m'}^{(b,c)} \, h_{\omega' \ell' m'}^{(b,c) \, *}(x) \bigg] \label{Kruskalexp} \quad .
\ee
    Here $\alpha^{(b,c)}$ and $\beta^{(b,c)}$ are Bogolubov coefficients, which can be computed using the usual scalar product. They are partially diagonal in the sense that $\alpha_{\omega \ell m \omega' \ell' m'} \sim \delta_{\ell,\ell'} \delta_{m,m'}$ and $\beta_{\omega \ell m \omega' \ell' m'} \sim (-1)^m \delta_{\ell, \ell'} \delta_{m,-m'}$ \cite{GoodAndersonEvans}. Therefore, for $\ell=0$, one has the simplification
\be
    p_{\omega 00}^{(b,c)}(x) = \int_0^\infty d\omega' \bigg[\alpha_{\omega \omega'}^{(b,c)} \, h_{\omega' 00}^{(b,c)}(x) + \beta_{\omega \omega'}^{(b,c)} \, h_{\omega' 00}^{(b,c) \, *}(x) \bigg] \label{Kruskalsimp} \quad ,
\ee
    with \cite{AndersonShohrehScofield}
\bes
    \begin{flalign}
        \alpha^{(b,c)}_{\omega \omega'} &= \frac{\sqrt{\omega'}}{2\pi \sqrt{\omega}}\left(\frac{1}{\kappa_{(b,c)}}\right)^{1+i\omega'/\kappa_{(b,c)}}(\epsilon-i\omega)^{i\omega'/\kappa_{(b,c)}}\, \,\Gamma\left(\delta-\frac{i\omega'}{\kappa_{(b,c)}}\right) \label{bogo1} \quad , \\
        \beta^{(b,c)}_{\omega \omega'} &= \frac{\sqrt{\omega'}}{2\pi \sqrt{\omega}}\left(\frac{1}{\kappa_{(b,c)}}\right)^{1-i\omega'/\kappa_{(b,c)}}(\epsilon-i\omega)^{-i\omega'/\kappa_{(b,c)}}\, \,\Gamma\left(\delta+\frac{i\omega'}{\kappa_{(b,c)}}\right) \label{bogo2} \quad .
    \end{flalign} \label{bogo}
\ees
    Here $\epsilon$ and $\delta$ are integrating factors such that $0<(\epsilon,\delta)\ll 1$. 
    
    For points in the static patch, the Boulware modes have the general form
\be
    h_{\omega' 00}^{(b,c)}(x) = \frac{Y_{00}}{r\sqrt{4\pi \omega'}}e^{-i \omega' t}\chi_{\omega' 0}^{(b,c)}(r) \label{Boulware2} \quad .
\ee
    The radial mode functions are solutions to the equation
\be
    \frac{d^2}{dr_*^2}\chi_{\omega' 0}^{(b,c)}(r) + \bigg[\omega'^{\, 2}-V_\textnormal{eff}(r)\bigg]\chi_{\omega' 0}^{(b,c)}(r) = 0 \label{radialeqn} \quad ,
\ee
    with
\be
    V_\textnormal{eff}(r) = \frac{f(r)}{r}\frac{d}{dr} f(r) \label{Veff} \quad .
\ee
    Note that $V_\mathrm{eff} \to 0$ on the black hole and cosmological horizons.

    To numerically solve $\eqref{radialeqn}$, two linearly independent solutions $\chi^\infty_{(R,L)}$ can be constructed which, for $r \to r_c$, have the following asymptotic behaviors
\bes
    \begin{flalign}
        \chi_R^\infty(r) &= e^{i \omega' r_*} \quad , \\
        \chi_L^\infty(r) &= e^{-i \omega' r_*} \quad .
    \end{flalign}
\ees
    Since $V_\mathrm{eff}=0$ on $H^-_b$, approaching the black hole horizon $r=r_b$ the solutions take the general form
\bes
    \begin{flalign}
         \chi^\infty_{R}(r) &= E_{R}(\omega')e^{i\omega' r_*}+F_{R}(\omega')e^{-i\omega' r_*} \quad , \\
         \chi^\infty_{L}(r) &= E_{L}(\omega')e^{i\omega' r_*}+F_{L}(\omega')e^{-i\omega' r_*} \quad ,
    \end{flalign}
\ees
    where $E_{(R,L)}$ and $F_{(R,L)}$ are scattering parameters that can be determined numerically. The radial contributions to the Boulware modes $\chi^{(b,c)}_{\omega' 0}$ are \cite{AndersonShohrehScofield}
\bes
    \begin{flalign}
        \chi^b_{\omega' 0}(r) &= \frac{\chi^\infty_R(r)}{E_R(\omega')} \quad , \\
        \chi^c_{\omega' 0}(r) &= \chi^\infty_L(r) - \frac{E_L(\omega')}{E_R(\omega')}\chi^\infty_R(r) \quad .
    \end{flalign} \label{chibc}
\ees

\section{Kruskal Modes and the Hadamard Function: \\ Numerical Results}

\begin{figure}[h]
    \centering
    \includegraphics[scale=0.3575]{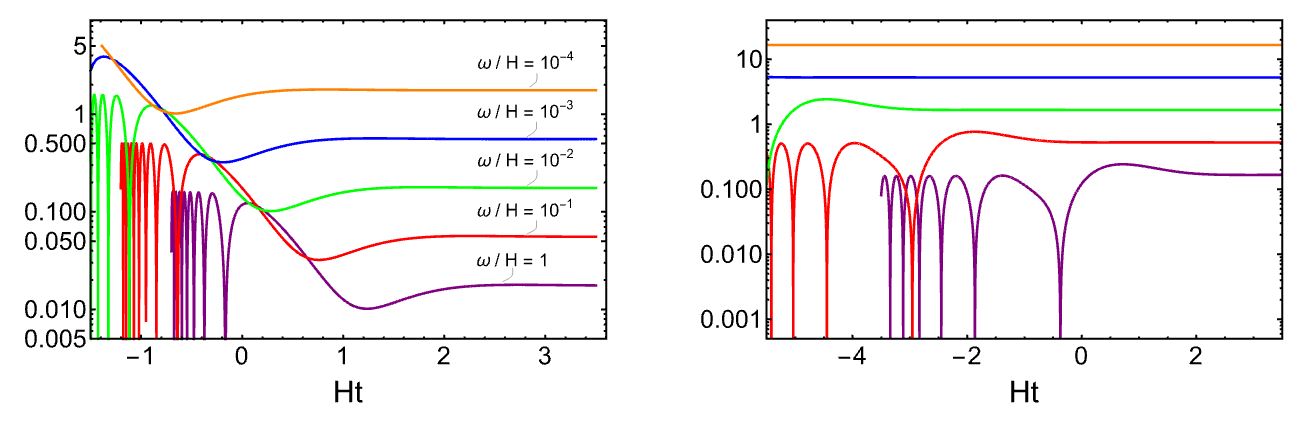}
    \caption{The late time behavior is shown for the magnitude of the real components of the black hole (left) and cosmological (right) Kruskal modes $H^{-1/2} \, p^{(b,c)}_{\omega 0 0}$, for a variety of frequencies in the range $10^{-4}\leq \omega / H \leq 1$.  Both plots are evaluated at a radial coordinate value $Hr=0.5$ between the horizons $Hr_b=0.1$ and $Hr_c\approx 0.95$. From highest to lowest at late times, the frequencies of the curves in the two plots are the same.} \label{Kruskalt}
\end{figure}

    The time evolution of the real components to the modes $p^{b}_{\omega 0 0}$ and $p^{c}_{\omega 0 0}$ are shown for various frequencies $\omega$ in Fig.~\ref{Kruskalt} for the choice $Hr=0.5$, which is between the horizons $Hr_b=0.1$ and $Hr_c\approx 0.95$. Note, for some frequencies the early-time oscillatory behavior of the plot curves has been artificially truncated to assist with visualization.
    
    It can be seen for all frequencies considered that at sufficiently late times the real parts of $p^b_{\omega 0 0}$ and $p^c_{\omega 0 0}$ approach nonzero constant values which decrease with frequency. The late time constant values approached by the magnitudes of the real parts of the $p^c_{\omega 0 0}$ modes are larger that those for $p^b_{\omega 00}$ and more spread out in terms of frequency. In both cases, the time required for a mode to achieve a constant value increases with frequency. As a given $p^b_{\omega 00}$ mode transitions from its early-time oscillatory behavior to its late-time approach to a constant value, there is a significant decrease in its magnitude. 
    The duration of this decrease appears to be roughly the same for all frequencies considered. In contrast, the $p^c_{\omega 00}$ modes approach late time constant values that are comparable to the amplitudes of their early time oscillations. Furthermore, it was found that the imaginary components of both $p^{(b,c)}_{\w 00}$ modes vanish at late times.

\begin{figure}[h]
    \centering
    \includegraphics[scale=0.33]{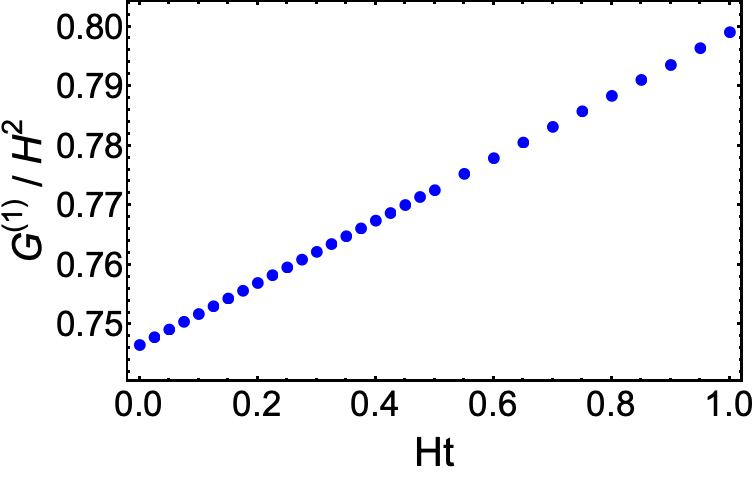}
    \caption{Numerical solutions for the $\ell=0$ contribution to the 4D Hadamard function $G^{(1)}/H^2$ at late times are shown as a function of the time $Ht$ for radially separated points $Hr=0.3$ and $Hr'=0.5$ between the horizons $H r_b = 0.1$ and $Hr_c \approx 0.95$.}
    \label{G1sample}
\end{figure}

    The $\ell=0$ contribution to the Hadamard function $\eqref{G1kruskal}$ is plotted as a function of the static time coordinate $t$ in Fig.~\ref{G1sample} for equal times and for radial points $Hr=0.3$ and $Hr'=0.5$, between the horizons $Hr_b=0.1$ and $Hr_c\approx 0.95$. There is clear linear growth over the times considered.
    
    To see the effect of position on the 4D Hadamard growth rate, multiple radial coordinate pairs in the static patch were investigated for fixed values of $r_b$ and $r_c$. To within two significant digits the growth rate was numerically determined to be $\frac{\partial}{\partial t} \, G^{(1)}(x,x') = 0.052 \, H^3$, providing strong evidence that not only does the linear instability persists in 4D, but the growth rate is independent of the choice of radial coordinates $r$ and $r^\prime$.

\section{Late-time Approximation to the Hadamard Function}

    In this section, an analytic approximation will be derived for the $\ell=0$ contribution to the Hadamard function growth rate at late times. First note that \eqref{Kruskalsimp}, with \eqref{bogo} and \eqref{Boulware2}, can be written as
\begin{flalign}
    p^{(b,c)}_{\w 00}(x) &= i \kappa_{(b,c)} \int_0^\infty d \w' \bigg\{ \left(\frac{A^{(b,c)}(r)}{\omega' + i \kappa_{(b,c)} \delta}\right) e^{-i \w'\left[ t - \kappa^{-1}_{(b,c)} \ln \left(\omega \, \kappa^{-1}_{(b,c)} \right) \right]} \nonumber \\
    & \qquad \qquad \qquad \qquad \qquad \qquad \qquad - \left(\frac{B^{(b,c)}(r)}{\omega' - i \kappa_{(b,c)} \delta}\right) 
    e^{i \w'\left[ t - \kappa^{-1}_{(b,c)} \ln \left(\omega \, \kappa^{-1}_{(b,c)}\right) \right]} \bigg\}  \quad , \label{pderiv} 
\end{flalign}
    with
\be
    A^{(b,c)}(r) = e^{- \pi \omega' \kappa^{-1}_{(b,c)}} B^{(b,c) *}(r) = \frac{e^{\frac{1}{2} \pi \w' \kappa^{-1}_{(b,c)}}}{8 \pi^2 \kappa_{(b,c)} \sqrt{\w}} \, \Gamma\left(1- \frac{i \w'}{\kappa_{(b,c)}}\right) \frac{\chi^{(b,c)}_{\w' 0}(r)}{r}  \quad .
\ee
    For fixed values of $\w$ and $r$, and at sufficiently late times, the integrand in~\eqref{pderiv} oscillates rapidly in $\w'$ due to the factors of $e^{\mp i \w' t}$. Therefore, the primary contributions to this integral originate from small values of $\w'$.  The reason these contributions, and therefore the integral, do not vanish in the limit $t \to \infty$ is that the integrand contains singularities at $\w' = 0$.

    In what follows, a late-time approximation for this integral is derived that, in agreement with the numerical results in Fig.~\ref{Kruskalt}, shows the modes $p^{(b,c)}_{\w 0 0}$ approach constant values at late times. The analytic approximation to the Hadamard function is derived by isolating the contribution from those Kruskal modes which have approximately constant values at any given time $t$.  

   To obtain the late-time approximations for $p^{(b,c)}_{\w 0 0}$ one can evaluate $A^{(b,c)}$ and $B^{(b,c)}$ at $\omega'=0$ with the result
\be
    A^{(b,c)}(r) = B^{(b,c)*}(r) = \frac{1}{8\pi^2 \kappa_{(b,c)} \sqrt{\w}} \frac{\chi^{(b,c)}_{0 0}(r)}{r} \quad . 
\ee    
    The calculation of $\chi_{\omega' 0}^{(b,c)}$ in this limit has been done for $\ell =0$ in~\cite{AndersonFabbriBalbinot}, where it was found that
\be
     \chi^\infty_R(r) = \chi^{\infty \, *}_L(r) = \frac{r}{r_c} \quad .
\ee
    The scattering parameters are
\begin{flalign}
        E_{R} = \frac{1}{2}\left( \frac{r_b^2 + r_c^2}{r_b \, r_c} \right) \quad , \quad E_L = \frac{1}{2}\left(\frac{r_b^2 - r_c^2}{r_b \, r_c} \right) \quad .
\end{flalign}
    Substituting these expressions into \eqref{chibc} gives 
\be 
    \frac{\chi^{(b,c)}_{0 0}(r)}{r} = \frac{ 2  \, r_{(b,c)}}{r_b^2 + r_c^2}  \label{BoulwareSmallfreq}\quad .
\ee
    This implies that $B^{(b,c)}=B^{(b,c)*}$ when $\omega'=0$. Therefore, one can take $\omega' \to -\omega'$ for the second integral in \eqref{pderiv} and combine it with the first integral to obtain the late-time approximation to the Kruskal modes
\be
    p^{(b,c)}_{\w 0 0}(x) \approx i \kappa_{(b,c)} A^{(b,c)}(r) \int_{-\infty}^\infty d\omega' \left( \frac{1}{\w' + i \kappa_{(b,c)} \delta} \right) e^{-i \w'\left[ t - \kappa^{-1}_{(b,c)} \ln \left(\omega \, \kappa^{-1}_{(b,c)}\right) \right]} \quad .
\ee
    This integral can be evaluated using contour integration. The result, which is valid for $\kappa_{(b,c)}\, t \gg \ln \left(\omega \, \kappa^{-1}_{(b,c)} \right)$, is
\be
    p^{(b,c)}_{\w 0 0}(x) \approx \frac{r_{(b,c)}}{2 \pi (r_b^2+ r_c^2) \sqrt{\omega}} \quad , \label{p0-approx}
\ee
    which indicates that the Kruskal modes approach constant values at late times.

    For $t' = t$, the contribution to $G^{(1)}(x,x')$ that grows linearly in time originates from those Kruskal modes which have already approached the above constant values. The contribution from these modes to the Hadamard function is obtained by substituting \eqref{p0-approx} into \eqref{G1kruskal} and inserting the upper limit cutoffs $\gamma \kappa_{(b,c)} e^{\kappa_{(b,c)} t}$ into the integrals for some $\gamma \ll 1$. Then the late-time approximation to the growth rate is
\be
    R_{4D} \equiv \frac{\partial}{\partial t} G^{(1)}(t,r,\theta,\phi;t,r',\theta',\phi') = \frac{1}{2\pi^2(r_b^2+r_c^2)^2}\big(r_b^2 \kappa_b + r_c^2 \kappa_c \big)  \label{Ranalytic} \quad .
\ee
    
\begin{figure}[h]
    \centering
    \includegraphics[scale=0.4]{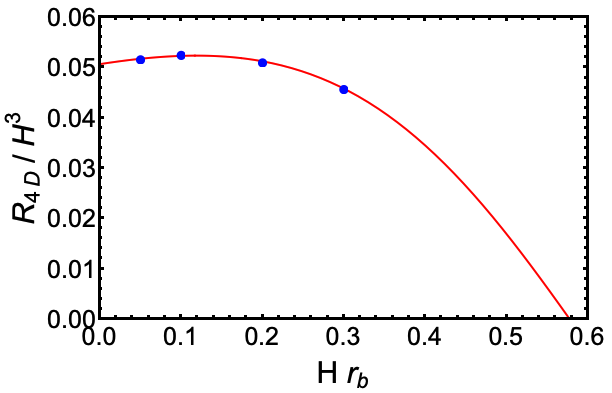}
    \caption{The analytic approximation to $R_{4D} / H^3$ is shown as a function of the black hole radius $H r_b$. The blue dots are results from the full numerical calculations of this rate.}
    \label{G12D4Dgrowth}
\end{figure}

    In Fig.~\ref{G12D4Dgrowth}, the analytic approximation for the 4D growth rate is plotted as a function of the black hole radius $r_b$ along with the numerically computed rate for several values of $r_b$. There is excellent agreement between the two~\footnote{Note that as $r_b$ increases, $r_c$ necessarily decreases, thereby causing the radial solutions $\chi^\infty_{(R,L)}$ between the horizons to exhibit more rapid, oscillatory behavior in $r$. Thus, it becomes increasingly difficult to numerically compute the 4D Hadamard growth rate for larger values of $r_b$.}.
        
    Interesting features emerge upon comparing the 2D and 4D cases. From $\eqref{R-2D}$ and $\eqref{kappas}$, it can be seen that as $r_b \to 0$, the 2D SdS growth rate diverges since $R_{\rm 2D} \sim 1/r_b$, while in 4D SdS the growth rate is $R_{4D}=H^3/2\pi^2$, which is in agreement with that found in \cite{AllenFolacci} for 4D de Sitter space in cosmological coordinates.  For intermediate values of the black hole horizon radius, there exists a maximum that occurs in 4D when $H r_b \approx 0.12$. There is no maximum in 2D.  Both the 2D and 4D Hadamard function growth rates terminate when $H r_b=H r_c = 1/\sqrt{3}$ since $\kappa_b$ and $\kappa_c$ vanish in this limit.

\section{Discussion and Conclusions}

    The spherically symmetric contribution to the Hadamard two-point function has been computed in 4D SdS for a massless minimally-coupled scalar field in the Unruh state for points that are separated along a coincident time hypersurface in the region between the black hole and cosmological horizons. Two sets of modes characterize the Unruh state. For one set, the modes are positive frequency with respect to the Kruskal time on the past black hole horizon and for the other they are  positive frequency with respect to the Kruskal time on the past cosmological horizon. The mode equation is not separable in either set of Kruskal coordinates, so Bogolubov transformations were used to express the Kruskal modes in terms of wave packets of Boulware modes. The Boulware modes were obtained using separation of variables and the equation for the radial part of the modes was solved numerically.

    It was found that the Kruskal modes approach nonzero constants at late times for fixed spatial points; the higher the frequency of a mode, the later in time that damping and subsequent approach to a nonzero constant occurs. Similar behavior manifests in 2D where there is no scattering, but the nonzero constant value approached is different than that in 4D.

    The contribution to the 4D Hadamard function from the spherically symmetric modes in the Unruh state is found to exhibit unbounded linear growth at late times in terms of the usual static time coordinate for radially separated points. This same effect was found previously in 2D \cite{AndersonTraschen}. However, the rate of growth is suppressed in 4D by scattering effects. In particular, the contribution from the Kruskal modes that are defined on the past black hole horizon is significantly smaller than that from the Kruskal modes that are defined on the past cosmological horizon.  
    
    These effects are corroborated by the derivation of an analytic expression for the rate of growth of the Hadamard function using various approximations. The analytic expression was found to be in good agreement with the numerical results, and furthermore, in the limit that the black hole event horizon vanishes, it agrees with the result found previously for 4D de Sitter spacetime \cite{AllenFolacci} in cosmological coordinates.

    There are infrared divergences associated with the Boulware modes which in 2D have been seen to have a significant impact not only on the late-time behaviors of the Kruskal modes, but also on the late-time behavior of the Hadamard function.  In particular, it was shown in~\cite{AndersonShohrehScofield} for 2D asymptotically flat static black holes, that when a delta function potential is included in the radial mode equation, scattering effects remove the infrared divergences in the Boulware modes.  In this case, the Kruskal modes vanish at late times for fixed spatial points and no linear growth in time of the Hadamard function occurs. In 4D SdS, scattering effects do not remove the infrared divergences for the spherically symmetric modes of the massless minimally-coupled scalar field. However, they are expected to remove these divergences for the modes with higher order spherical harmonics. Thus, the only contribution to the Hadamard function that is expected to result in unbounded linear growth is from the spherically symmetric modes.

    The linear growth in time of the Hadamard function is a sign of some type of instability associated with the Unruh state. In 2D Schwarzschild spacetime, the linear growth in time of the Hadamard function leads to a linear growth in time of the quantity $\la \phi^2 \ra$~\cite{and-peak-gho}. There is good reason to expect this will also occur for 4D SdS. However, the stress-energy tensor is not expected to undergo any linear growth in time because it involves two derivatives of the Hadamard function.
    
    The effect found here for SdS is similar in nature to the well-known linear growth in time of the Hadamard function in 4D de Sitter spacetime for the Bunch-Davies state shown in~\cite{AllenFolacci}. There it was found that for de Sitter space there are alternative homogeneous and isotropic vacuum states for which no such instability occurs.  The stress-energy tensor for the massless minimally-coupled scalar field in this case has an asymptotic de Sitter-invariant value for these states that is different than that for the Bunch-Davies state, which is also de Sitter invariant.    
    
    What distinguishes SdS from de Sitter space is the presence of an eternal black hole. The Unruh state is thought to be the most physically relevant state for an eternal black hole because it yields a flux of radiation emanating from the black hole equivalent to that predicted by Hawking for black holes that form from collapse at late times. As mentioned above, it is clear from our result is that for a massless minimally-coupled scalar field in 4D SdS, there is some type of instability for the Unruh state.  One can ask whether such an instability is likely to persist in models in which the black hole forms from collapse, since then the initial vacuum state for the field is not the Unruh state. This has been tested in 2D in the case of a Schwarzschild black hole that forms from the implosion of a null shell of radiation~\cite{AndersonTraschen}. In this case, there is no past horizon and there is a well-defined initial vacuum state. For that state, the leading order contribution to the Hadamard function at late times has the same linear growth in time as for an eternal Schwarzschild black hole in 2D. Therefore, it is quite possible that the instability found here for the Unruh state in 4D SdS would persist in models in which a black hole in de Sitter space forms from collapse.

\newpage

\section*{Acknowledgments}

    P. R. A. would like to thank Jennie Traschen and Shohreh Gholizadeh Siahmazgi for helpful conversations. P.R.A. would also like to thank Northwestern University's Center for Interdisciplinary Exploration and Research in Astrophysics (CIERA) for hosting him during a visit in August of 2023. P.R.A. was supported, in part, by the National Science Foundation under grants No. PHY-1912584 and PHY-2309186 to Wake Forest University. S. P. was supported by the Leverhulme Trust under grant No. RPG2021-299.

\end{document}